\documentclass[a4paper,twocolumn]{Gaia2004}

\usepackage{times}      
\usepackage{epsfig}     
\usepackage{natbib}     

\title{Introducing adapted Nelder \& Mead's downhill simplex method to a fully automated analysis of eclipsing binaries}

\author{A.~Pr\v sa, T.~Zwitter}
\affil{University of Ljubljana, Dept.~of Astrophysics, Jadranska 19, 1000 Ljubljana, Slovenia}

\bibpunct{(}{)}{;}{a}{}{,}  

\begin{document}


\maketitle

\begin{abstract}
  
Eclipsing binaries are extremely attractive objects because absolute physical parameters (masses, luminosities, radii) of both components may be determined from observations. Since most efforts to extract these parameters were based on dedicated observing programs, existing modeling code is based on interactivity. Gaia will make a revolutionary advance in shear number of observed eclipsing binaries and new methods for automatic handling must be introduced and thoroughly tested. This paper focuses on Nelder \& Mead's downhill simplex method applied to a synthetically created test binary as it will be observed by Gaia.

\end{abstract}

\section{Introduction}

Many assessments have already been done for Gaia expected harvest of eclipsing binaries (EB) to V$<$15, where both photometric and RV observations will have been available (see Munari et al.~2001, Zwitter et al.~2003, Marrese et al.~2004 and others). Out of 50 million observed stars, roughly 100\,000 will be double-lined eclipsing binaries. However, based on experience from Hipparcos, out of 1 billion stars observed to V$<$20, there will be $\sim$\emph{2 million} (Eyer, these proceedings) eclipsing binaries without spectroscopic observations, but with quite decent photometric accuracy. This study presents a new approach developed for automatic reduction of observed data along with an estimate of how much we may expect to obtain from them.

\section{The method}

Obtaining physical parameters from observations is an inverse problem solved numerically by a modeling program. Affirmed and most widely used is the WD code \citep{wilson1971}, which features Differential Corrections method (DC) powered by the Method of Multiple Subsets (MMS) \citep{wilson1993}. DC method has already been applied successfully to automatic parameter extraction (e.g.~Wyithe \& Wilson 2001, Wyithe \& Wilson 2002, Pr\v sa 2003 and others), but its original philosophy is based on interactive monitoring of each convergence step. The algorithm is very fast and works well if the discrepancy between the observed and computed curves is relatively small, but it tends to diverge or give physically unmeaningful results if the discrepancy is large. As part of an effort to create a reliable and powerful package for EB analysis, a complementing minimization scheme is proposed.

\subsection{Nelder \& Mead's Downhill Simplex}

There are two main deficiences of the DC method that are especially striking. {\bf 1)} Once a DC method converges to a minimum, there is no way of telling whether that minimum is local or global; even if it is local, the method is stuck and cannot escape. {\bf 2)} The main source of divergence and the loss of accuracy in the DC algorithm is the computation of numerical derivatives. 

To circumvent these two problems, Nelder \& Mead's downhill Simplex\footnote{Nelder \& Mead's downhill simplex is not to be confused with linear or non-linear programming algorithms, which are also referred to as Simplex methods.} method \citep{nelder1965}, hereafter NMS, is implemented. Since NMS doesn't compute derivatives but relies only on function evaluations, it is a promising candidate for our purpose. Basic form of the NMS method along with WD implementation was first proposed by \cite{kallrath1987}. We take a step further and adapt the method specifically to EBs.

NMS method acts in $n$-dimensional parameter hyperspace. It constructs $n$ vectors $\mathbf p_i$ from the vector of initial parameter values $\mathbf x$ and the vector of step-sizes $\mathbf s$ as follows:
\begin{equation}
\mathbf p_i = (x_0, x_1, \dots, x_i + s_i, \dots, x_n)
\end{equation}

These vectors form $(n+1)$ vertices of an $n$-dimensional simplex. During each iteration step, the algorithm tries to improve parameter vectors $\mathbf p_i$ by modifying the vertex with the highest function value by simple geometrical transformations: reflection, reflection followed by expansion, contraction and multiple contraction \citep{galassi2003}. Using these transformations, the simplex moves through parameter space towards the deepest minimum, where it contracts itself.

PHOEBE ({\rm http://phoebe.fiz.uni-lj.si}) is a software package built on top of the WD code that extends its basic functionality to encompass, among other extensions summarized in \citet{prsa2005}, the NMS method. It is especially suited for EBs: powered by heuristic scans, parameter kicking and conditional constraining, the method is able to efficiently escape from local minima.

\subsection{Heuristic Scan}

NMS is a robust method that always converges, but it can converge to a local minimum, particularly since parameter hyperspace in vicinity of the global minimum is typically very flat, with lots of local minima. In adition, global minimum may be shadowed by data noise and degeneracy.

Heuristic scan is a method by which a minimization algorithm selects a set of starting points in parameter hyperspace and starts the minimization from each such point. It then sorts all solutions by the cost function (the $\chi^2$, for example) and calculates parameter histograms and convergence tracers for given hyperspace cross-sections (specific examples are given in Section \ref{results}). The way of how the algorithm selects starting points is determined by the user: the points may be gridded, stochastically dispersed, distributed according to some probability distribution function (PDF) etc. The basic idea of heuristic scan is to obtain decent statistics of adjusted parameter values from which a fair and realistic error estimate may be given.

\subsection{Parameter Kicking} \label{parameter_kicking}

Another possible approach to detect and escape from local minima is to use some stochastic method like Simulated Annealing (SA). However, such methods are notoriously slow. Since the EB hyperspace is typically very flat, stochastic methods would be practical only in the vicinity of the global minimum. Thus instead of full-featured SA scan, a simple new procedure has been developed that achieves the same effect as stochastic methods, but in significantly shorter time.

The idea is as follows: whenever NMS reaches a minimum within a given accuracy, the algorithm runs a globality assessment on that minimum. If we presume that standard deviations $\sigma_k$ of observations are estimated properly and that they apply to all data points regardless of phase or nightly variations, we may use them for $\chi^2$ weighting:
\begin{equation}
\chi_k^2 = \sum_{i=1}^M w_k (x_i - \bar x)^2 = \frac 1{\sigma_k^2} \sum_{i=1}^M (x_i - \bar x)^2,
\end{equation}
where index $i$ runs over $M$ measurements within a single data-set and index $k$ runs over $N$ data-sets (different photometric curves). Since the variance is given by:
\begin{equation}
s_k^2 = \frac{1}{N_k-1} \sum_i (x_i - \bar x)^2,
\end{equation}
we may readily express $\chi_k^2$ as:
\begin{equation}
\chi_k^2 = (N_k - 1) \frac{s_k^2}{\sigma_k^2}.
\end{equation}
and the overall $\chi^2$ value as:
\begin{equation}
\chi^2 = \sum_k (N_k - 1) \left( \frac {s_k}{\sigma_k} \right)^2.
\end{equation}

If $\sigma_k$ are fair and all data-sets contain approximately the same number of observations, then the ratio $s_k / \sigma_k$ is of the order unity and $\chi^2$ of the order $N M$. This we use for parametrizing $\chi^2$ values:
\begin{equation} \label{eq_lambda}
\chi_0^2 = N M, \quad \lambda := \left( \chi^2 / \chi_0^2 \right) \,: \textrm{quantization.}
\end{equation}

Parameter kicking is a way of knocking the obtained parameter-set out of the minimum: using the Gaussian PDF, the method randomly picks an offset for each parameter. The strength of the kick is determined by the Gaussian dispersion $\sigma_{\mathrm{kick}}$, which depends on $\lambda$: if the value is high, then the kick should be strong, but if it is low, i.e.~around $\lambda \sim 1$, then only subtle perturbations should be allowed. Experience shows that a simple expression such as:
\begin{equation}
\sigma_{\mathrm{kick}} = \frac{0.5 \lambda}{100}
\end{equation}
works very reliably. This causes $\sigma_{\mathrm{kick}}$ to assume a value of $0.5$ for $10 \sigma$ offsets and $0.005$ for $1 \sigma$ offsets, being linear in between. Note that this $\sigma_{\mathrm{kick}}$ is \emph{relative}, i.e.~given by:
\begin{equation}
\sigma_{\mathrm{kick}}^{\mathrm{abs}} = x \, \sigma_{\mathrm{kick}}^{\mathrm{rel}},
\end{equation}
where $x$ is the value of the given parameter.

\subsection{Conditional Constraining}

Having purely photometric (LC) observations, it is impossible to determine absolute physical parameters of the observed binary. However, if the distance to EB is measured independently or if additional assumptions about the EB are set, even purely photometric observations can yield absolute values of physical parameters. If additional constraints are imposed on the model from the outside, the model is referred to as \emph{conditionally constrained} (CC'd).

Two CCs are immediately evident: {\bf 1)} astrometric measurements on-board Gaia may be used to measure the distance with the accuracy of $\sim$11\,$\mu$as at V=15 to $\sim$165\,$\mu$as at V=20 (\citet{redbook}4); {\bf 2)} since a substantial number of stars are main sequence stars, M--L, L--T and T--R relations may be adopted as constraints. See \citet{prsa2005} for details on CC implementation in PHOEBE.

\section{Simulation} \label{simulation}

To test the suitability of NMS for EBs, we built a partially-eclipsing synthetic main-sequence F8\,V--G1\,V binary using PHOEBE. Table \ref{params} lists all of its principal parameters. The simulation presented in this paper is based exclusively on photometric data: two sets of observations corresponding to Johnson B and V filters are created, each with 82 points with Poissonian scatter $\sigma_{\mathrm{obs}} = 0.02$mag at $V=20$, values typical to expect from Gaia (\citet{redbook}4).

Simulation flow is the following: all physical parameters set for adjustment ($a$, $i$, $T_1$, $T_2$, $\Omega_1$, $\Omega_2$, $L_1$ and $L_2$) were displaced by $\sim$50\%. The obtained set was used as initial guess for the NMS. In the first part of the simulation the method converges to a solution using only heuristic scan and parameter kicking, which yields \emph{relative} values of parameters. In the second part the simulation was additionally CC'd by the main-sequence constraint to obtain \emph{absolute} values of parameters.

\begin{table}[!ht]
\caption{Principal parameters of the simulated binary.}
\label{params}
\begin{center}
\leavevmode
\begin{tabular}{lrrr}
\hline \hline
Parameter [units] & & Binary & \\
& F8\,V & & G1\,V \\
\hline \\
$P_0$ [days]                       & &  1.000 & \\
$a\,\,\,[R_\odot]$                 & &  5.524 & \\
$q=m_2/m_1$                        & &  0.831 & \\
$i\,\,\,[{}^\circ]$                & & 85.000 & \\
$T_\mathrm{eff}\,\,\,[\mathrm K]$  &   6200 & &   5860 \\
$L\,\,\,[L_\odot]$                 &  2.100 & &  1.100 \\
$M\,\,\,[M_\odot]$                 &  1.236 & &  1.028 \\
$R\,\,\,[R_\odot]$                 &  1.259 & &  1.020 \\
$\Omega\,\,\,[-]\,{}^{\mathrm{(a)}}$ &  5.244 & &  5.599 \\
\hline \\
\multicolumn{4}{p{6.4cm}}{${}^{\mathrm{(a)}}$~Unitless effective potentials defined by \citet{wilson1979}.} \\
\end{tabular}
\end{center}
\end{table}

\section{Results} \label{results}

Heuristic scan over all adjustable parameters has been performed to obtain accurate convergence statistics: over 2\,000 uniformly distributed starting points in parameter hyperspace were used during simulation. We present the results of overall minimization step-by-step.

\begin{figure}[!t]
\begin{center}
\leavevmode
\centerline{\epsfig{file=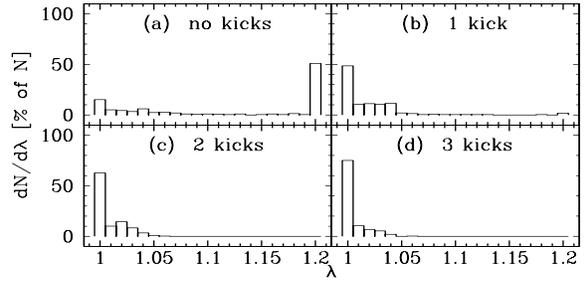,width=1.0\linewidth,height=0.5\linewidth}}
\end{center}
\caption{$\lambda$-histograms for 3 consecutive parameter kicks. Top-left figure demonstrates how numerous are local minima and how difficult it is for NMS to circumvent them. Other three figures show significant improvement by using parameter kicking. Histograms consist of 20 bins (single bin width is 0.01). The last bin encompasses all higher values of $\lambda$. Labels \,\emph{(a)}\, through \,\emph{(d)}\, are used consistently throughout the paper.}
\label{lambda_hist}
\end{figure}

{\bf a)\, Convergence assessment.} Depending on the bumpiness of the hyperspace, heuristic scan will generally yield different solutions from different starting points; it is our hope that only few of these solutions will account for most scans. To evaluate their quality, the globality assessment mechanism introduced in section \ref{parameter_kicking} is used to sort solutions by the depth of the reached minimum. Tests show that the NMS method itself is all-too-often stuck in local minima and only $\sim$15\% of all runs end up within one percent of ideal $\lambda$ ($\lambda$=1 in case of Fig.~\ref{lambda_hist}). However, tests also show that parameter kicking \emph{significantly} improves this percentage ($\sim$50\% after the first kick, $\sim$63\% after the second and $\sim$75\% after the third kick); even more, parameter kicking also enhances convergence speed after each kick.

\begin{figure}[!t]
\begin{center}
\leavevmode
\centerline{\epsfig{file=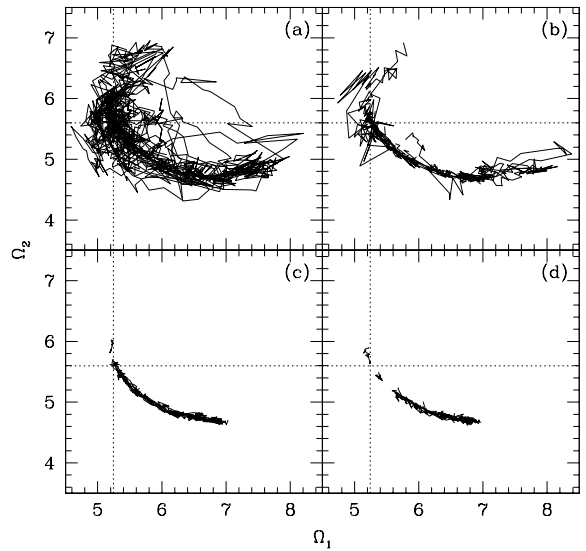,width=1.0\linewidth}}
\end{center}
\caption{Convergence tracers for 3 consecutive parameter kicks. These plots trace convergence steps within 2D cross-sections of the hyperspace, revealing areas of minima and degeneracy. Cross-hairs mark the location of the global minimum.}
\label{tracers}
\end{figure}

Although the values of $\lambda$ may seem promising, they don't necessarily guarantee that the corresponding solution is optimal. Rather, additional assessments should be done. Fig.~\ref{tracers} shows \emph{convergence tracers}: 2D cross-sections of parameter hyperspace tracing convergence from each starting point of heuristic scan to the corresponding solution. Such tracers clearly show areas of minima and degeneracy. Since the location of the global minimum is not known in real life, extra care should be taken to \emph{never} blindly trust the statistics of such a degenerate problem.

{\bf b)\, Statistics of obtained parameters.}

\begin{figure}[!t]
\begin{center}
\leavevmode
\centerline{\epsfig{file=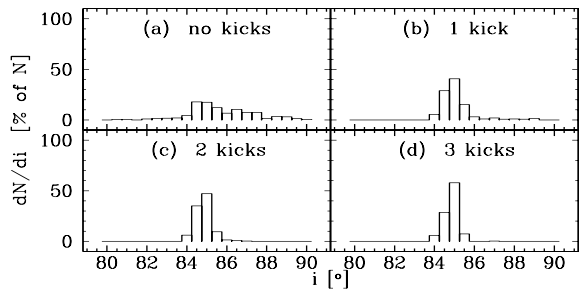,width=1.0\linewidth,height=0.5\linewidth}}
\end{center}
\caption{Histogram of the inclination for 3 consecutive parameter kicks. Histogram consists of 20 bins with $0.5^\circ$ each. The solution is symmetric to $i=90^\circ$, but we adopt $i<90^\circ$ by convention.}
\label{ihist}
\end{figure}

The usual practice in literature, when listing obtained parameters from the model, is to give their values with \emph{formal} errors, i.e.~standard deviations reported by the used numerical method. These errors are often too optimistic, since degeneracy and noise noticeably contribute to the overall error. NMS powered by heuristic scan and parameter kicking has the advantage of obtaining parameter errors \emph{statistically}, independent of the method itself. Fig.~\ref{ihist} shows an example of obtained histogram for inclination $i$. It is evident that for the NMS without parameter kicking (and similarly for {\bf any} other numerical method that cannot escape from local minima) any eclipsing system is a tie; for a fully minimized solution (bottom right plot on Fig.~\ref{ihist}) the error is simply standard deviation of the Gaussian being fitted over the histogram, yielding roughly $0.5^\circ$.

{\bf c)\, Conditional constraining.}

\begin{figure}[!t]
\begin{center}
\leavevmode
\centerline{\epsfig{file=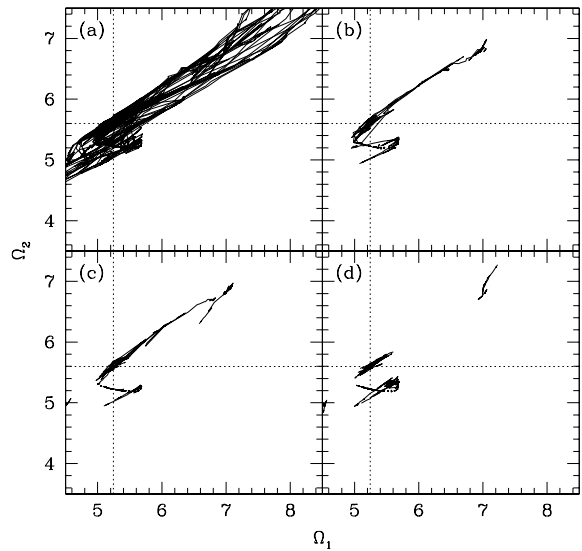,width=1.0\linewidth}}
\end{center}
\caption{Main-sequence constrained convergence tracers for 3 consecutive parameter kicks. Comparing this result to Fig.~\ref{tracers} clearly shows that the intersection of both areas indeed gives the right solution. This is expected, since our synthetic binary is in fact a main-sequence binary.}
\label{tracers_ms}
\end{figure}

Inclination is the only intrinsic parameter that may be obtained in absolute sense from photometric observations. Using CC this deficiency is removed: any particular CC adds one or more implicit parameter ties into the system. It basically introduces an intersection plane with the otherwise degenerate part of the hyperspace, thus eliminating degeneracy. Fig.~\ref{tracers_ms} demonstrates how a main-sequence constraint breaks the degeneracy for gravity potentials $\Omega_{1,2}$.

\section{Discussion} \label{conclusion}

The idea behind the NMS implementation is \emph{not} to replace the DC method but to \emph{complement} it. DC is created for interactive usage and converges in discrete steps that need monitoring. NMS on the other hand aims to automate this process so that intermediate monitoring is no longer necessary, which is a key goal for Gaia. DC is one of the fastest methods (WD's {\tt DC} in particular, since it is optimized for EBs), but may easily diverge. On expense of speed, NMS is one of the most robust algorithms for solving non-linear problems and never diverges. Finally, both DC and NMS methods suffer from degeneracy and may be stuck in local minima. To overcome this, DC is complemented by the MMS and NMS is complemented by heuristic scan and parameter kicking. These differences in intent make a combination of both methods a powerful engine for solving the inverse problem.

\end{document}